\documentclass{article}

\usepackage{PRIMEarxiv}

\usepackage[utf8]{inputenc} 
\usepackage[T1]{fontenc}    
\usepackage{hyperref}       
\usepackage{url}            
\usepackage{booktabs}       
\usepackage{amsfonts}       
\usepackage{nicefrac}       
\usepackage{microtype}      
\usepackage{lipsum}
\usepackage{fancyhdr}       
\usepackage{graphicx}       
\usepackage{amsmath}
\usepackage{amsfonts}
\graphicspath{{media/}}     

\title{Effects of Mutual Coupling on {Degree of Freedom} and Antenna Efficiency in Holographic MIMO Communications}

\author{Shuai S. A. Yuan $^{1}$, Xiaoming Chen $^{2}$, Chongwen Huang $^{1}$ and Wei E. I. Sha $^{1,}$*\\ \\
$^{1}$ College of Information Science and Electronic Engineering, Zhejiang University, Hangzhou 310027, China.\\
$^{2}$ School of Information and Communications Engineering, Xi'an Jiaotong University, Xi'an 710049, China.\\ \\
\texttt{Correspondence: weisha@zju.edu.cn}}

\begin{document}
\maketitle
\begin{abstract}
The holographic multiple-input-multiple-output (MIMO) communications refer to the MIMO systems built with ultra-dense antenna arrays, whose channel models and potential applications have attracted increasing attentions recently. When the spacing between adjacent array elements is larger than half wavelength, the effect of mutual coupling can generally be neglected in current antenna designs. However, in holographic MIMO communications, the influence of strong mutual coupling on antenna characteristics is inevitable, resulting in distorted radiation patterns and low radiation efficiencies. In this paper, starting from the analytical correlation and efficiency models, we investigate how the mutual coupling affects the capacity of a space-constrained MIMO system from the aspects of degree of freedom (DOF) and antenna efficiency. The involved fundamental concepts of correlation, DOF, efficiency and mutual coupling are crucial for both antenna and wireless-communication engineers when designing emerging MIMO communication systems.
\end{abstract}


\section{Introduction}
The multiple-input-multiple-output (MIMO) technology for modern wireless communications has achieved great success over the past two decades \cite{telatar1999capacity, tse2005fundamentals}. MIMO technology can provide both the degree of freedom (DOF) and power gains for enhancing channel capacity \cite{Massa2022, Cui2020}. As the spectrum resource becomes scarce for fifth-generation (5G) and beyond 5G applications, it is essential to explore strategies for enhancing the channel capacity without occupying larger physical space. Consequently, more and more antennas are tightly arranged into a constrained area, which leads to the massive MIMO \cite{TL2014, SL2018}, holographic MIMO \cite{Chongwen2022,Pizzo2020}, and continuous aperture MIMO \cite{sayeed2010continuous,zhang2021continuous} communications. Recently, the channel models and potential applications of  holographic MIMO systems are frequently discussed \cite{Wei2022,Pizzo2022}. In such dense arrays, the effect of mutual coupling is inevitable and should be carefully taken into account.

The capacity of a space-constrained MIMO system is dependent on both the DOF and signal-to-noise ratio (SNR) \cite{tse2005fundamentals}. DOF refers to the rank of the correlation matrix of a MIMO system, which characterizes the spatial-multiplexing performance. The DOF limit of a space-constrained MIMO system has been discussed from the perspectives of both the information theory \cite{signal2005space,muharemovic2008antenna,Loyka2004} and electromagnetic (EM) theory \cite{piestun2000electromagnetic,MD2019,pierri1998information,Bucci1989}, proving that the DOF is fundamentally bounded by the aperture size of array in arbitrary propagating environments.  The total SNR is usually fixed as a specific value when evaluating the capacity. Both the DOF and SNR of a MIMO system are affected by mutual coupling, since the distorted embedded radiation patterns will change the correlations between antennas, and the coupling of power will reduce radiation efficiency (related to SNR). 

Regarding the effect of mutual coupling on the channel capacity of a MIMO system, it will generally reduce the embedded radiation efficiencies and may lower the correlation between antennas \cite{xiaoming2018,Wallace2004,Kulandai2008,Masouros2013,Chen2016,Sawaya2006,RJ2002,Nie2004}, then finally decrease the MIMO capacity \cite{Kildal2004}. {Recently, various decoupling technologies have been developed for improving phased array and MIMO performances \cite{Wu2014,Cui2012,Chu2013,You2016,Kishk2017,Chen2015,Li2022}.} With measured or simulated embedded radiation patterns and efficiencies, several numerical methods are proposed for conveniently estimating the performance of a practical MIMO system rather than testing in a reverberant chamber, such as the Kronecker model \cite{xiaoming2013} and the ray-tracing method \cite{RT2016}. Moreover, the theoretical bounds of MIMO systems are discussed, including the Hannan's limit for efficiency \cite{Hannan1964,Kildal2016} and the DOF limit in rich-scattering (Rayleigh) environments \cite{Pizzo2020,ShuaiPra}. Nevertheless, a clear clarification of the physical foundations of correlation, DOF, and efficiency is still necessary. Also, the role that mutual coupling plays in holographic MIMO communications is not well investigated.

In this paper, mainly two contributions are made. First, based on a one-dimensional (1-D) microstrip dipole array, we investigate how the mutual coupling affects the capacity of holographic MIMO systems from the aspects of DOF and efficiency. The analytical DOF, efficiency and capacity models are discussed in detail and then compared to the numerical results. Second, Hannan's efficiency limit is introduced for accurately characterizing the decrease of antenna efficiency brought by mutual coupling, which can be easily embedded in the channel model of holographic MIMO communications. {Rather than simply calculating the MIMO capacity in terms of efficiency and correlation matrix, we are focusing on analyzing the DOF and efficiency limit of an aperture-constrained array through both analytical models and full-wave simulations, which is particularly important to the emerging holographic MIMO communications. The proposed theoretical methods and presented numerical results could provide physical understanding and engineering guideline for designing and evaluating next-generation MIMO communications.}
\section{MUTUAL COUPLING IN MIMO ANTENNA ARRAY}
When one antenna in a MIMO array is excited, the input power will be coupled from the excited antenna to the others due to free-space radiation, conduction current, and surface wave. The principle of mutual coupling has been well discussed in the literature \cite{balanis2015antenna, kildal2015foundations}, and we are not going to revisit them in detail here, but give some visualized demonstrations of the effects of mutual coupling. In this paper, a microstrip printed dipole with reflecting board is used for investigating mutual coupling, as demonstrated in Fig. 1 (a-b). The upgraded versions of this kind of antenna are widely used for commercial basestation applications \cite{li2009equivalent,gou2014compact}. In order to arrange sufficient antennas along one direction, a 1-D array with fixed length and uniform element spacing is employed, as seen Fig. 1 (c). 
\begin{figure}[ht!]
	\centering
	\includegraphics[width=3.4in]{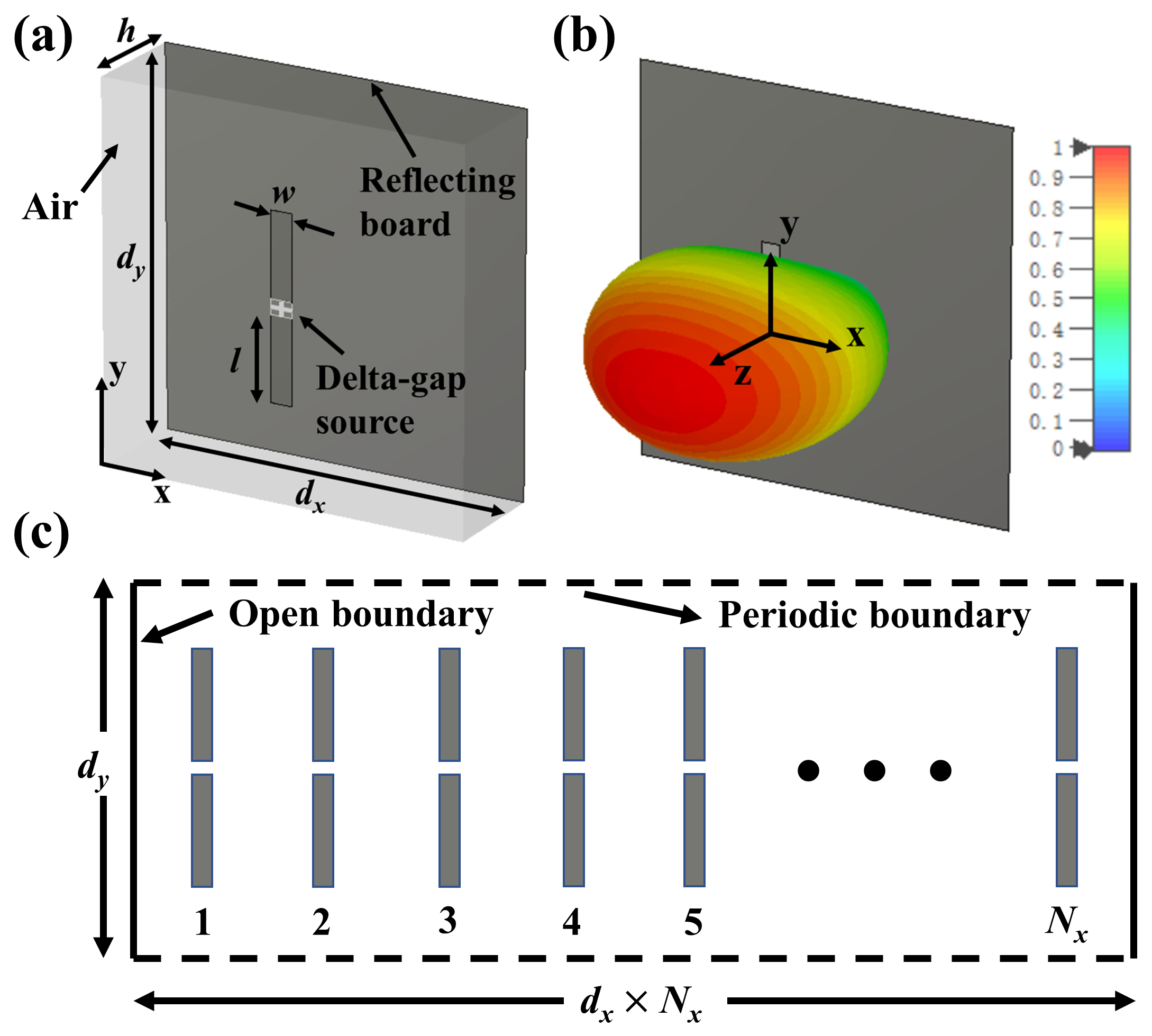}
	\caption{Configuration of the microstrip dipole array. (a) Array element working at 5 GHz (radiation efficiency is 0.95, directivity is 5), with the wavelength $\lambda_0=60$ mm. $w=3$ mm,  $l=22.5$ mm, $h=15$ mm, and $d_x=d_y=60$ mm. (b) Radiation pattern of the isolated array element (in linear scale). (c) Structure of the one-dimensional (1-D) antenna array. $d_y=60$ mm, and the total length along the $x-$axis is fixed as $L_x$. The open boundary condition is applied along the $x-$axis and the periodic boundary condition is applied along the $y-$axis.}
	\label{MIMO}
\end{figure}
\noindent
Based on this configuration, the embedded radiation patterns and efficiencies of a four-element 1-D array at 0.3 $\lambda_0$ ($\lambda_0$ is the free-space wavelength) element spacing are depicted in Fig. 2. {The embedded radiation pattern of one antenna in array can be acquired by feeding the antenna while making all the other antennas well matched \cite{Pozar2003, Kildal2016}, which can be obtained by computational electromagnetic methods.} It can be observed that the radiation patterns are severely distorted and the efficiencies are largely decreased. These are typical effects brought by mutual coupling, as out-of-phase currents are induced on the surfaces of surrounding antennas when exciting one antenna. {In large-size arrays, the embedded radiation patterns of the antennas in the central area will not be distorted too much due to the simultaneous influences of the surrounding antennas \cite{kildal2015foundations}, and only the antennas at the edge will be severely distorted. Therefore, the effect of mutual coupling on the DOF of a large-size array is negligible, but the effect will become significant for a small-size and dense array.}

\begin{figure*}[ht!]
	\centering
	\includegraphics[width=6in]{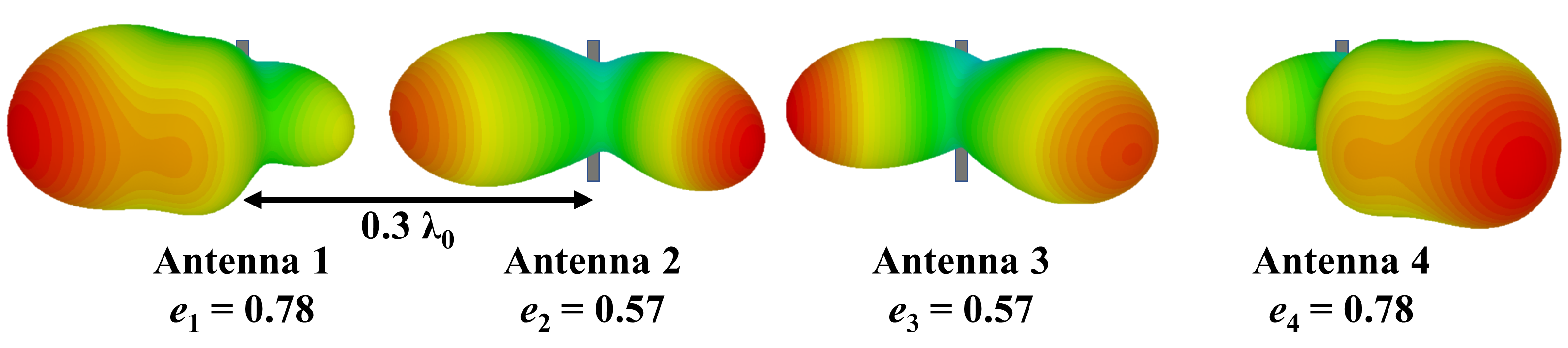}
	\caption{Embedded radiation patterns and efficiencies $e_n$ of a four-element array, the element spacing is 0.3 $\lambda_0$.}
	\label{MIMO}
\end{figure*}

\section{ANALYTICAL CORRELATION, DOF AND EFFICIENCY MODELS}
\subsection{CORRELATION MODEL}
\begin{figure*}[ht!]
	\centering
	\includegraphics[width=6in]{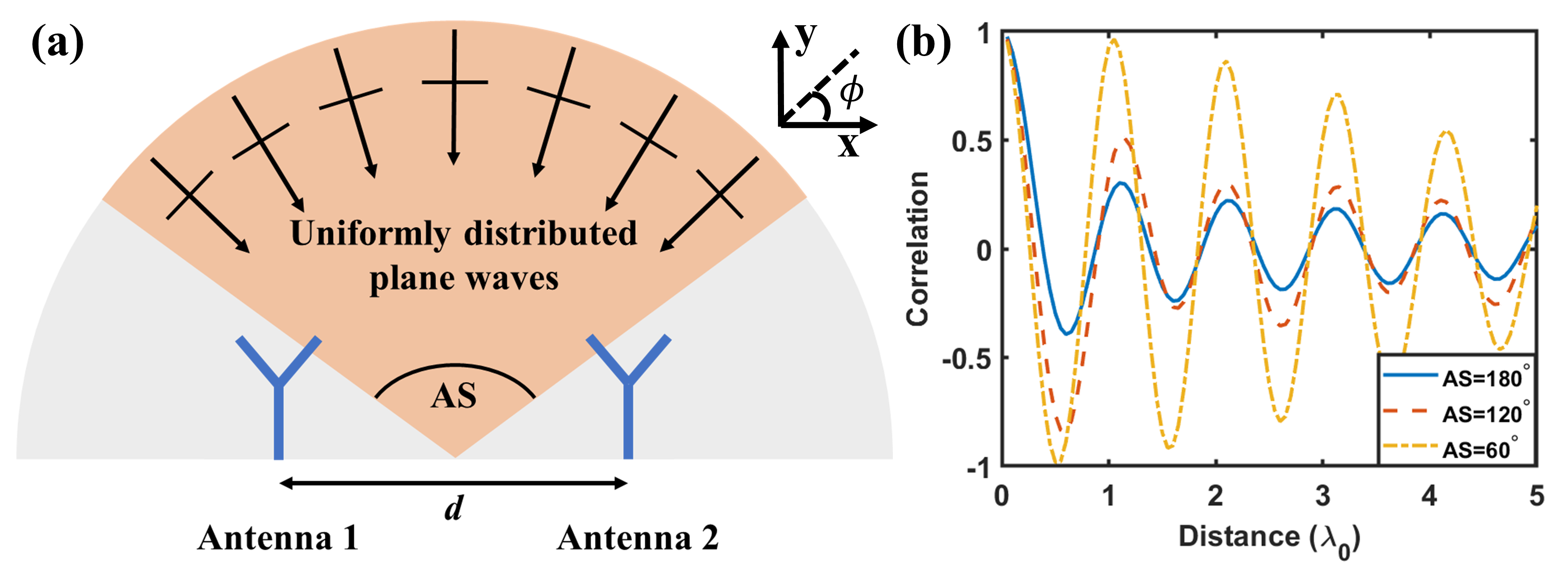} 
	\caption{Antenna correlation model in 2-D multipath environment. (a) Correlation between two antennas, the coming waves are modeled as uniformly distributed plane waves within a specific angular spread (AS). (b) 2-D correlation functions under different ASs (AS = 180$^{\circ}$ for Clarke's model).}
	\label{MIMO}
\end{figure*}
For an $M\times N$ antenna array, the correlation between each pair of antennas can be written as an $MN\times MN$ correlation matrix $\Phi$. The capacity of a practical MIMO system can be precisely evaluated with $\Phi$ and an efficiency matrix. DOF, which characterizes the spatial multiplexing performance of a MIMO system, is defined as the number of significant eigenvalues of $\Phi$. Physically, the eigenvalues of $\Phi$ represent the strengths of available spatial EM modes, and the corresponding eigenvectors tell us how to excite these modes. Since DOF is completely determined by the correlations between antennas, it is particularly important to mathematically express the correlation models and physically understand how antenna radiation characteristics (incorporating mutual coupling) influence these models.

One of the most commonly-used correlation models is the Clarke's model in two-dimensional (2-D) multipath environment \cite{clarke1968statistical,imoize2021standard}, which describes the correlation between antennas as a function of distance. In Clarke's model, the antennas are regarded as point receivers, and the receiving waves are modeled as TM-polarized ($E_z$) far-field plane waves uniformly coming from a half space (angular spread is equal to 180$^{\circ}$) or full space, as depicted in Fig. 3 (a). The total signal received by one antenna can be regarded as the superposition of these received plane waves. Therefore, we can easily formulate the correlation function between two antennas as (with sufficient number of plane waves)  \cite{Jakes}
\begin{equation}
\begin{aligned}
R_{2D}(d)&= \left\langle E_{z}(x,y)E_{z}^{*}(x+d,y)\right\rangle _{\rm{av}}\\
&=\frac{1}{{K}}\sum_{n=1}^{K} \left\langle \exp \left\{jk_0  \cos \phi_{n}d\right\}\right\rangle _{\rm{av}}\\
&=J_0(k_0d),
\end{aligned}
\end{equation}
where $K$ is the number of plane waves, $\phi_n$=$\pi n/K$ denotes the directions of plane waves, and $k_0$ is the free-space wavenumber. Moreover,  $^{*}$ is the conjugate operator, $\left\langle\cdot\right\rangle_{\rm{av}}$ represents the ensemble average, $d$ is the distance between the two receivers, and $J_0$ is the zero-order Bessel function. This correlation function is in fact the average of phase delays between the two antennas along different directions within the half-space angular spread. Within a smaller angular spread, the correlation will become larger, leading to a worse MIMO performance, see Fig. 3 (b). When the angular spread approaches 0, the correlation will be close to 1, which is equivalent to the line-of-sight (LOS) scenario in free space. 

\begin{figure*}[ht!]
	\centering
	\includegraphics[width=6in]{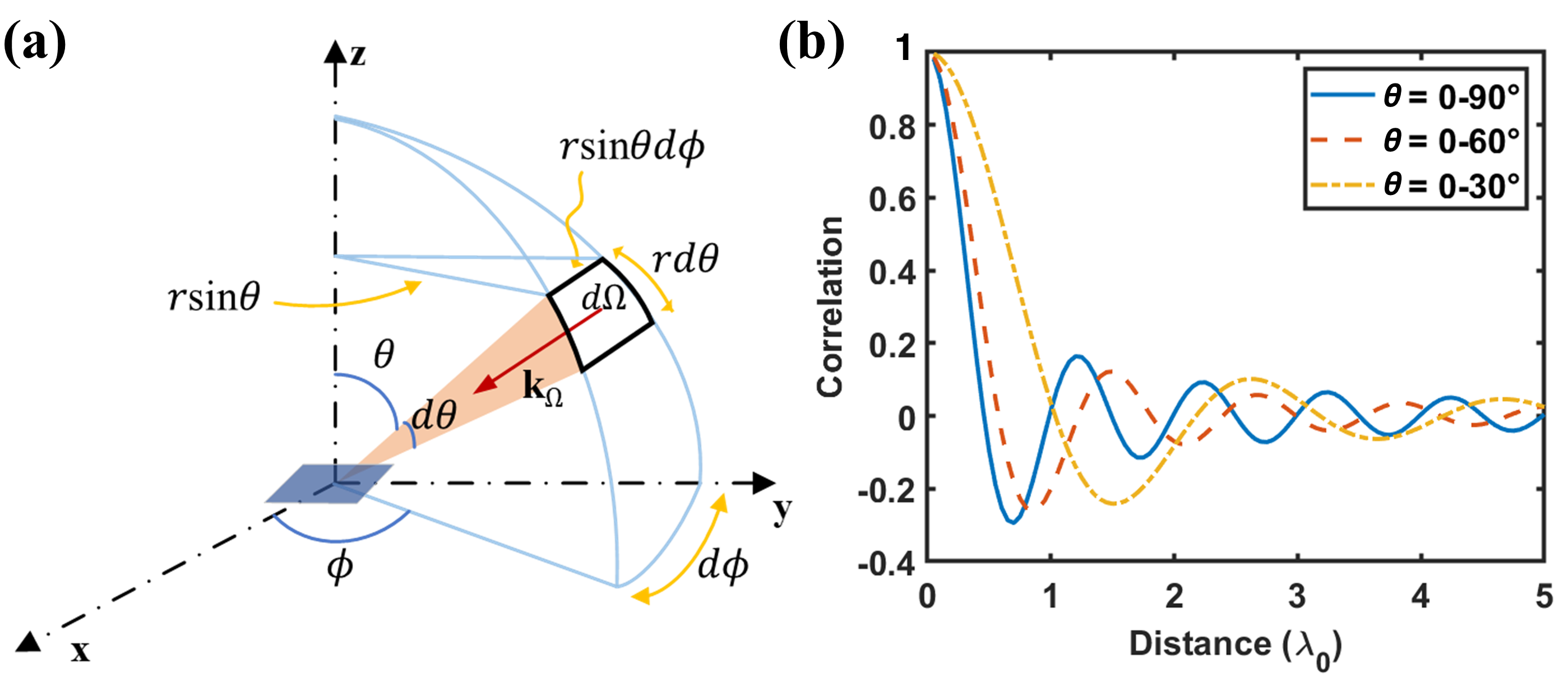}
	\caption{Antenna correlation model in 3-D multipath environments. (a) Illustration of the differential of solid angle ($d\Omega$), the received signal from the $\mathbf{k}_{\Omega}-$direction plane wave would be $\exp(j\mathbf{k}_{\Omega}\cdot\mathbf{r}_n)d\Omega$, where $\mathbf{r}_n$ denotes the position of the receiving antenna. (b) 3-D correlation functions under different ranges of $\theta$ .}
	\label{MIMO}
\end{figure*}
In three-dimensional (3-D) multipath environment, the basic principle of the correlation model is similar, the receiving waves are also modeled as uniformly distributed plane waves. However, the strengths of the received signals along different $\theta$ are different, since the received signal from the $\mathbf{k}_{\Omega}-$direction plane wave would be $\exp(j\mathbf{k}_{\Omega}\cdot\mathbf{r}_n)d\Omega$, where $\mathbf{r}_n$ denotes the position of the receiving antenna, as demonstrated in Fig. 4 (a). Hence, the superposed plane waves should multiply the weight of $\sin\theta$, which is the same as the integral over solid angle $\Omega$ in calculating envelope correlation coefficient (ECC) between antennas \cite{Andersen1987}. When the angular spread is half space ($\theta=0-90^{\circ}, \phi=0-360^{\circ}$), the expression for 3-D correlation function is close to $R_{3D}=sinc(k_0d)$ after normalizations \cite{pollock2003introducing}. The 3-D correlation functions under different ranges of $\theta$ are depicted in Fig. 4 (b), the null points of correlations are shifted under different angular ranges, which are quite different from the 2-D cases in Fig. 3 (b). In practice, the signal received by one antenna is simultaneously determined by its embedded radiation pattern (taking into mutual coupling) and angular spread, which will be further discussed in Section IV.
\subsection{DOF MODEL}
\begin{figure}[ht!]
	\centering
	\includegraphics[width=3.4in]{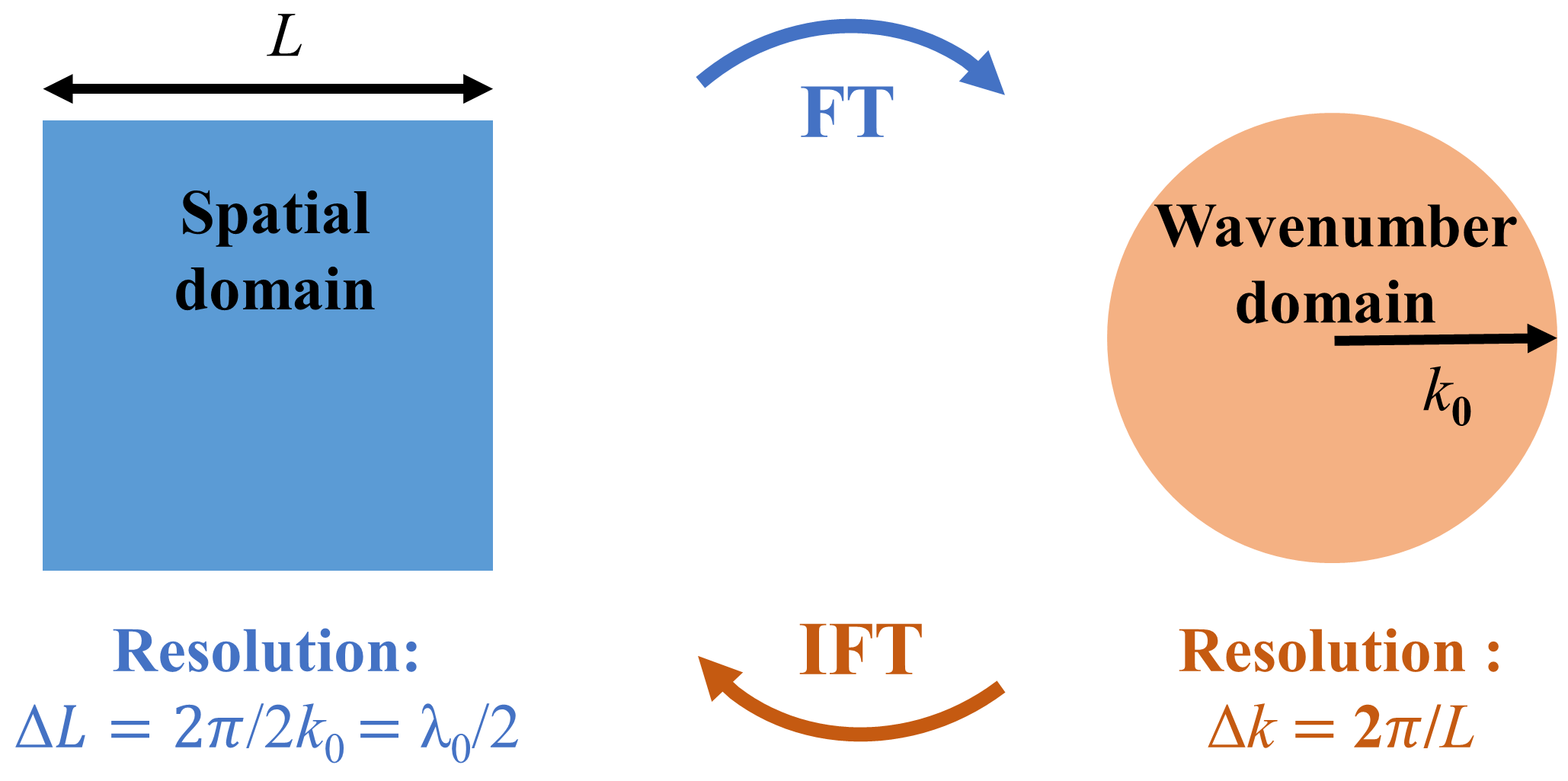}
	\caption{Theoretical resolution limits in spatial and wavenumber domains. Transformation from near field (spatial domain) to far field (wavenumber domain) is essentially a Fourier transform (FT), and the reverse process is an inverse Fourier transform (IFT). The wavenumber area (red circle) is defined by $k_x ^2+k_y ^2<k_0^2$ for propagating waves.}
	\label{MIMO}
\end{figure}
As depicted in Fig. 5, the physical size of a planar antenna array is $L \times L$, which can be regarded as an equivalent band-limited source function $\mathbf{J}(\mathbf{r}^{\prime})$. The far-field radiation pattern of this source can be calculated with a dyadic Green's function as
\begin{equation}\label{farfield}
\begin{aligned}
\mathbf{E}_{far}(\mathbf{k})=-j \omega \mu_{0} \frac{e^{-j k_0 r}}{4 \pi r} \int_{v}&\left(\mathbf{a}_{\theta} \mathbf{a}_{\theta}+\mathbf{a}_{\varphi} \mathbf{a}_{\varphi}\right) \cdot \mathbf{J}\left(\mathbf{r}^{\prime}\right)\\ &\exp \left(j \mathbf{k}\cdot \mathbf{r}^{\prime}\right) d \mathbf{r}^{\prime},
\end{aligned}
\end{equation}
\noindent where $\mathbf{a}_{\theta}\mathbf{a}_{\theta}$ and $\mathbf{a}_{\varphi}\mathbf{a}_{\varphi}$ represent the $\theta$ and $\varphi$ components of unit dyadic, and $ \mathbf{k}$ is the wave vector in free space. Obviously, the transformation from near-field (spatial domain) to far-field (wavenumber domain) is a Fourier transform (FT) , and the reverse process is an inverse Fourier transform (IFT). Hence, the minimum resolutions along one direction in wavenumber and spatial domains would be $\Delta k = 2\pi /L$ and $\Delta L=2\pi /2k_0=\lambda_0/2$, respectively. {When the angular spread of the incoming wave is complicated (non-isotropic scattering environment), the largest length (or area) in angular domain will become smaller, indicating that the resolution of spatial domain will become worse according to the FT, and the element spacing to reach the DOF limit will only be larger than $\lambda_0/2$.} Therefore, the $\lambda_0/2$ element spacing is generally sufficient for capturing all the information brought by EM waves coming from far field (although the embedded radiation patterns and angular spread are complicated in practice), unless the available wavenumber domain is expanded, like radiative near-field communications based on evanescent waves \cite{Shuai2021}. 

Furthermore, we can easily deduce the DOF limit of a MIMO system in isotropic (half space) multipath environment. According to Fig. 5, the resolution area for one EM mode in wavenumber domain would be $(\Delta k )^2$, and the total available wavenumber area is defined by
\begin{equation}\label{k_resource_of_half_space}
k_x ^2+k_y ^2<k_0^2,
\end{equation}
then the number of independent scattering channels (DOF limit) can be estimated by a division between the wavenumber area (red circle in Fig. 5) and the minimum resolution area $(\Delta k )^2$, which yields
\begin{equation}\label{upper_limit}
\mathrm{DOF} \leq   \frac{\pi k_0^2}{(\Delta k)^2}=\frac{\pi L^2}{\lambda_0^2}.
\end{equation}
\subsection{EFFICIENCY MODEL}
Radiation efficiency is a vital factor in evaluating capacity of MIMO antenna arrays, which may not be considered in some wireless communication models. To deduce the efficiency model, one should be aware that the maximum gain of an array element in a phased array is bound by its physical size according to Hannan's limit \cite{Hannan1964}. It has been found that Hannan's limit can well describe the decrease of efficiency brought by mutual coupling at small element spacing \cite{Kildal2016}. {We assume that the shape of array element is rectangular,} with the length of $d_x$ along the $x-$axis and the length of $d_y$ along the $y-$axis. The maximum available gain of such an array element in a phased array is $G_{e m b}=4\pi d_{x}d_{y}/\lambda_0^2$ \cite{Hannan1964}. At small element spacing, the radiation pattern of antenna would appear as a $\cos\theta$ shape. As a result, the directivity is the integral of the radiated power over $\Omega$, which yields the directivity $D_{e m b}\approx4$ ($\approx$ 6 dBi). Finally, the upper bound of the embedded antenna efficiency becomes
\begin{equation}
e_{e m b}=\frac{G_{e m b}}{D_{e m b}}=\frac{\pi d_{x}d_{y}}{\lambda_0^2 }.
\end{equation}
\noindent In consequence, when the antenna number is increasing in a space-constrained MIMO system, the reduction of element size will lead to the decrease of antenna efficiency characterized by (5). We will show in the following numerical examples that (5) is quite accurate compared with practical antenna efficiencies obtained from full-wave simulations. Interestingly, it can be found that the efficiency limit given by (5) is the same as the DOF limit given by (4) \cite{ShuaiPra}, indicating that the performances of MIMO communications are fundamentally bounded by the physical sizes of both array and element.
\section{PRACTICAL MIMO PERFORMANCES}
In this paper, we evaluate the practical performance of of MIMO antenna arrays under the Vertical-Bell-Labs-Space-Time (V-BLAST) architecture \cite{tse2005fundamentals}, where the transmitting side is ideal (the antennas are uncorrelated and their efficiencies are 1). The channel matrix is unknown to the transmitters, and the transmitting power is equally allocated. This architecture allows us to focus on the element spacing and antenna characteristics at the receiving side. {Assuming that there are $N_{t}$ transmitting antennas and $N_{r}$ receiving antennas ($N_{t}=N_{r}$),} the ergodic capacity incorporating receiving antenna effects can be written as \cite{xiaoming2013} \cite{Chen2016}
\begin{equation}
\begin{aligned}
C=E\left\{\log _{2}\left[\operatorname{det}\left(\mathbf{I}+\frac{\gamma}{N_{t}} \mathbf{R} \mathbf{H}_{w} \mathbf{H}_{w}^{H}\right)\right]\right\},
\end{aligned}
\end{equation}
where $E$ represents the mathematical expectation, $^H$ is the Hermitian operator, and $\mathbf{I}$ is the identity matrix. Moreover, $\gamma$ is the fixed total signal-to-noise ratio (SNR), the entries of $\mathbf{H}_{w}$ (matrix dimension is $N_{r} \times N_{t}$) are independent and identically distributed (i.i.d.) complex Gaussian variables denoting spatially white MIMO channel. {One should pay special attention to the normalization of $\mathbf{H}_w$ when evaluating the MIMO systems built with aperture-constrained arrays. {Usually, the $\mathbf{H}_w$ is normalized by making $\|{\mathbf{H}_w}\|_F^{2}=N_{t} N_{r}$ for fair comparisons \cite{tse2005fundamentals, Loyka2009}, where the $N_{t}N_{r}$ represents the array gain (or power gain) of the MIMO channel.\footnote{This is a frequently-used assumption in MIMO communications, the antennas are simplified as isotropic point sources, and the element spacing is an integer multiple of $\lambda_0/2$. Therefore, the array gain is simply equal to the number of antennas (assuming the antennas are lossless) .} However, the receiving array gain of an aperture-constrained array will not further increase when the element spacing becomes smaller than $\lambda_0/2$ \cite{Loyka2009} (assuming that the array gain is equal to the number of antennas before reaching the $\lambda_0/2$ spacing). Hence, the $\mathbf{H}_w$ should be normalized by 
		\begin{equation}
		\|{\mathbf{H}_w}\|_F^{2}=\left\{\begin{array}{lr}
		N_{t} N_{r}, & \text { element spacing} > \lambda_0/2 \\
		N_{t} N_{\lambda_0/2}, &  \text {element spacing} \le \lambda_0/2\\
		\end{array}\right.
		\end{equation}		
		where $N_{\lambda_0/2}$ is the number of receiving antennas at $\lambda_0/2$ element spacing and  $\|\cdot\|_F$ denotes the Frobenius norm.} The covariance matrix $\mathbf{R}$ is constructed by the entry-wise product between the correlation matrix $\boldsymbol{\Phi}$ and the embedded efficiency matrix $\boldsymbol{\Xi}$
	\begin{equation}
	\mathbf{R}=  \boldsymbol{\Phi}\circ\boldsymbol{\Xi}.
	\end{equation}
	The correlation matrix incorporating mutual coupling is \cite{xiaoming2017}
	\begin{equation}
	\boldsymbol{\Phi}=\left[\begin{array}{cccc}
	1 & \rho_{12} & \cdots & \rho_{1 N_{r}} \\
	\rho_{12}^{*} & 1 & \cdots & \rho_{2 N_{r}} \\
	\vdots & \vdots & \ddots & \vdots \\
	\rho_{1 N_{r}}^{*} & \rho_{2 N_{r}}^{*} & \cdots & 1
	\end{array}\right],
	\end{equation}
	where
	\begin{equation}
	\rho_{mn}=\frac{\oint G_{mn}(\Omega) \mathrm{d} \Omega}{\sqrt{\oint G_{mm}(\Omega) \mathrm{d} \Omega} \sqrt{\oint G_{nn}(\Omega) \mathrm{d} \Omega}},
	\end{equation}
	with
	\begin{equation}
	G_{m n}(\Omega)=\kappa E_{\theta m}(\Omega) E_{\theta n}^{*}(\Omega) P_{\theta}(\Omega)+E_{\phi m}(\Omega) E_{\phi n}^{*}(\Omega) P_{\phi}(\Omega),
	\end{equation}
	$ E_{\theta}(\Omega)$ and $ E_{\phi}(\Omega)$ are the $\theta-$ and $\phi-$polarized embedded radiation patterns, $\Omega$ represents the solid angle, ${P}(\Omega)$ is the angular power spectrum, and $\kappa$ is the cross-polarization discrimination (XPD). We assume the XPD is 1, and  ${P}(\Omega)=1$ over all the solid angles, which represents an isotropic-scattering polarization-balanced multipath environment. The embedded efficiency matrix $\boldsymbol{\Xi}$ is expressed by 
	\begin{equation}
	\Xi=\sqrt{\mathbf{e}}\sqrt{\mathbf{e}}^{T},
	\end{equation}
	with
	\begin{equation}
	\mathbf{e}=\left[e_{e m b 1}, e_{e m b 2},  \cdots, e_{e m b N_{r}}\right]^{T},
	\end{equation}
	where the embedded radiation efficiency of the $n$th antenna is calculated by the $S$ parameters assuming negligible ohmic loss
	\begin{equation}
	e_{e m bn}=1-\left|S_{1n}\right|^{2}-\left|S_{2n}\right|^{2}-\cdots -\left|S_{N_rn}\right|^{2}.
	\end{equation}
	It is worthy of mentioning that the embedded radiation efficiency is in fact deduced from the average of active $S$ parameters over all the solid angles \cite{Hannan1964, Kildal2016, Pozar2003}. 
	\section{NUMERICAL RESULTS}
	{Numerical investigations of the effects of mutual coupling are demonstrated based on the 1-D antenna array in Fig. 1(c), and the antennas are uniformly distributed along the $x-$axis within a fixed length $L_x=d_x\times N_x= 2 \lambda_0$, $N_x$ is the number of receiving antennas along the $x-$axis.} Through full-wave simulations empowered by commercial software CST MWS, embedded radiation patterns and efficiencies of the elements are extracted for evaluating the practical performances of the whole MIMO arrays, 
	\begin{figure}[ht!]
		\centering
		\includegraphics[width=3in]{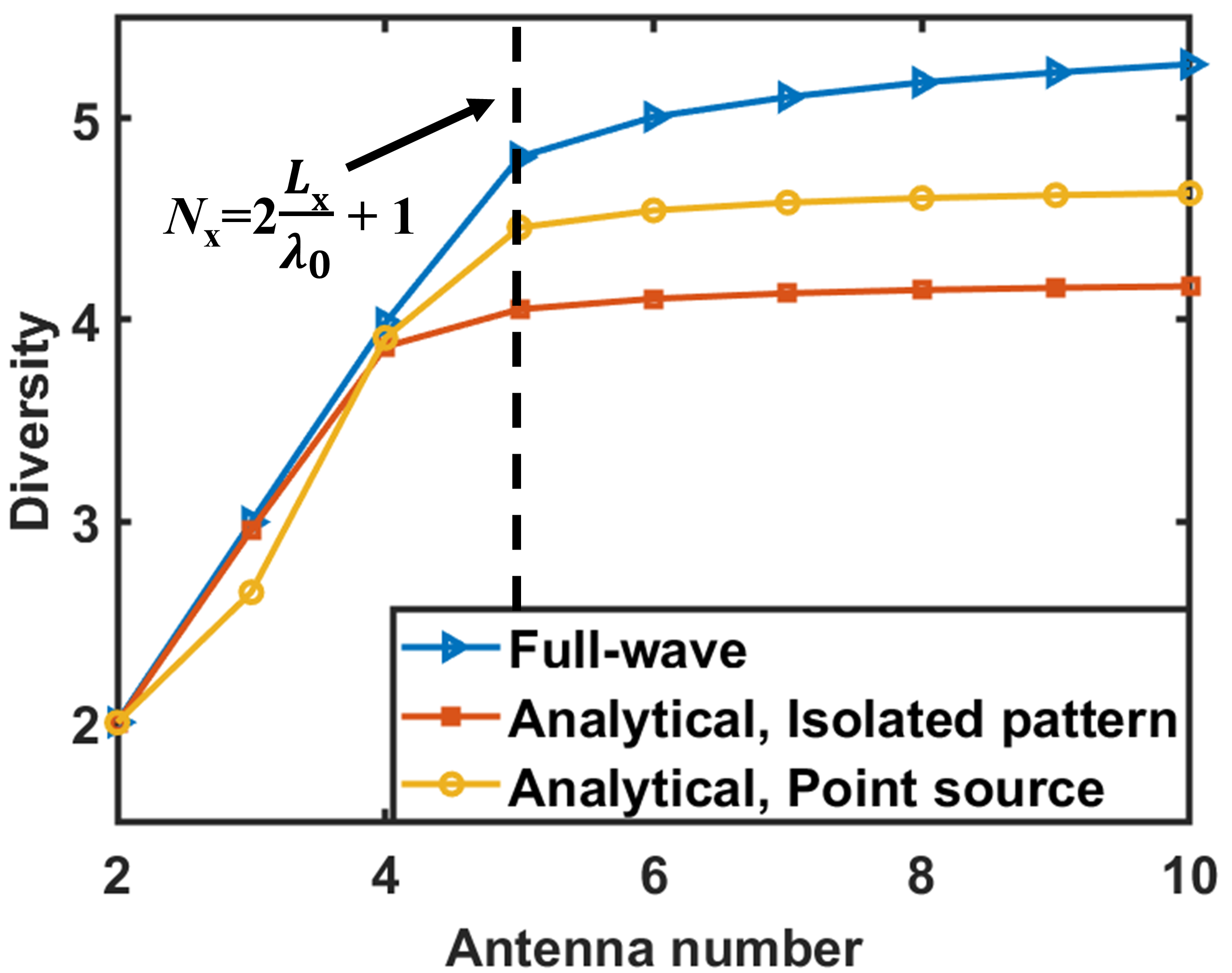}
		\caption{{Diversities of the receiving arrays in different scenarios.}}
		\label{MIMO}
	\end{figure}
	which are then compared to the analytical models and limits. For the correlation model without mutual coupling, the embedded radiation pattern of each antenna is the same as the radiation pattern of a single isolated antenna, but with an additional phase delay over each solid angle. If we write the radiation pattern of the isolated antenna as $ E^{iso}(\Omega)$, the embedded radiation pattern of the $n$th antenna would be
	\begin{equation}
	E(\Omega)=E^{iso}(\Omega)\exp(j\mathbf{k}_\Omega \cdot \mathbf{r}_n),
	\end{equation}
	where $\mathbf{k}_\Omega$ is the wave vector along the $\Omega$ direction, and $\mathbf{r}_n$ is the position of the $n$th antenna. Besides, the DOF performance is conveniently characterized by the diversity instead of counting significant eigenvalues \cite{NJ2005}. The diversity of a MIMO array can be calculated by
	\begin{equation}
	\Psi\left(\boldsymbol{\Phi}\right)=\left(\frac{\operatorname{tr}\left(\boldsymbol{\Phi}\right)}{\left\|\boldsymbol{\Phi}\right\|_{F}}\right)^{2}=\frac{\left(\sum_{i} \sigma_{i}\right)^{2}}{\sum_{i} \sigma_{i}^{2}},
	\end{equation}
	where $\operatorname{tr}(\cdot)$ represents the trace operator and $\sigma_{i}$ is the $i$th eigenvalue of the correlation matrix $\boldsymbol{\Phi}$. At the receiving side, the diversity represents the equivalent number of uncorrelated antennas of the receiving array.
	\subsection{EFFECT OF MUTUAL COUPLING ON DOF}
	{The DOF performances in different scenarios are characterized by the diversities demonstrated in Fig. 6.} At this array size ($L_x=2 \lambda_0$), it can be observed that the mutual coupling will slightly enhance DOF at small element spacing, as the deformations of embedded radiation patterns may lead to lower correlations between antennas \cite{xiaoming2020}. With increasing antenna number, the tendencies of the diversities with and without mutual coupling are almost the same, i.e., the DOF cannot significantly increase when the antenna number is larger than $2L_x/\lambda_0+1$. Therefore, even for holographic MIMO communications with strong mutual coupling, ignoring the mutual coupling will not bring large errors in the analysis of DOF. {Meanwhile, the effect of mutual coupling on DOF will become negligible when the array size becomes larger \cite{kildal2015foundations}.} 
	\begin{figure}[ht!]
		\centering
		\includegraphics[width=3in]{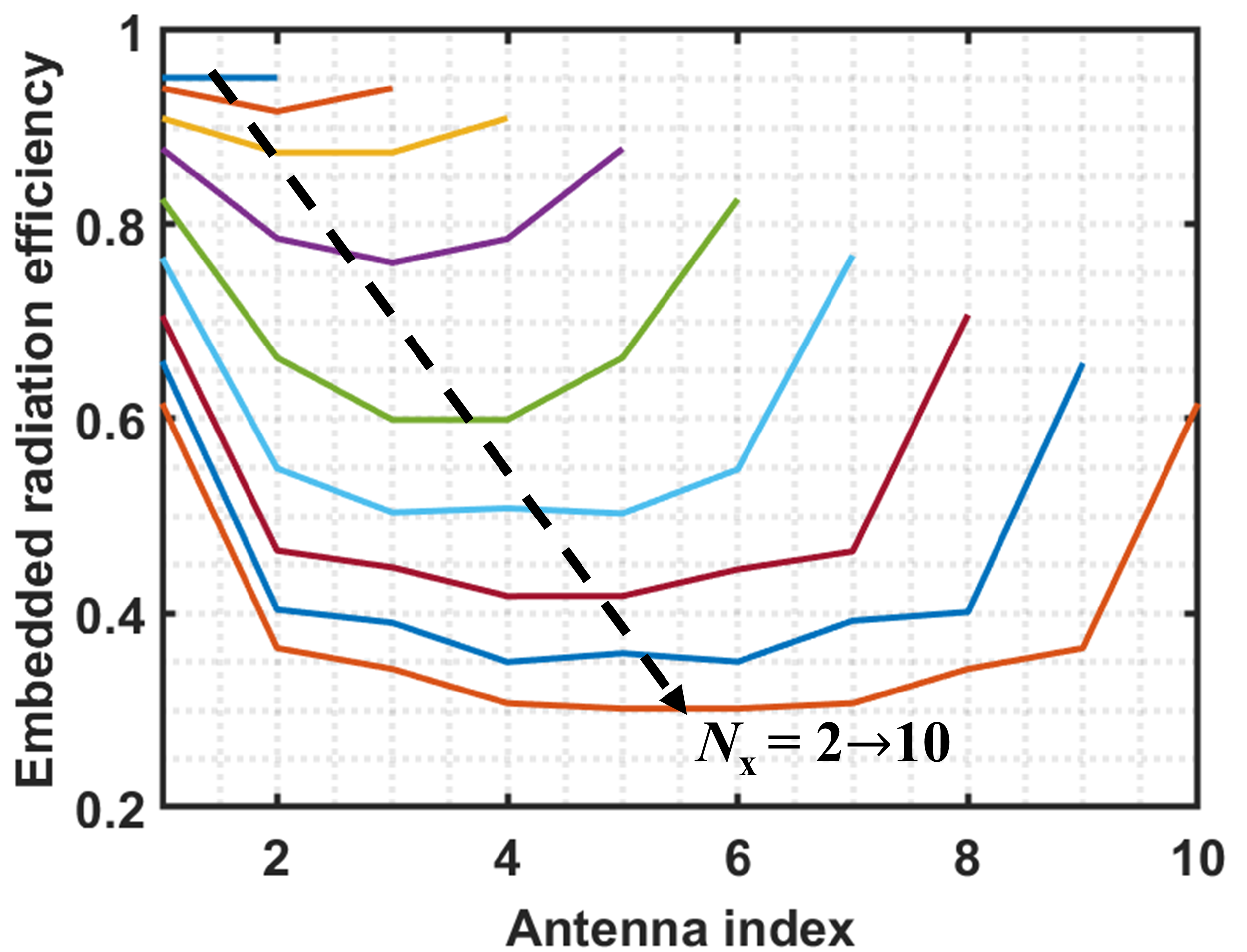}
		\caption{{Embedded radiation efficiencies of the array elements obtained from full-wave simulations, where the lengths of solid lines along the $x-$axis denote the numbers of antennas.}}
		\label{MIMO}
	\end{figure}
	\begin{figure}[ht!]
		\centering
		\includegraphics[width=3in]{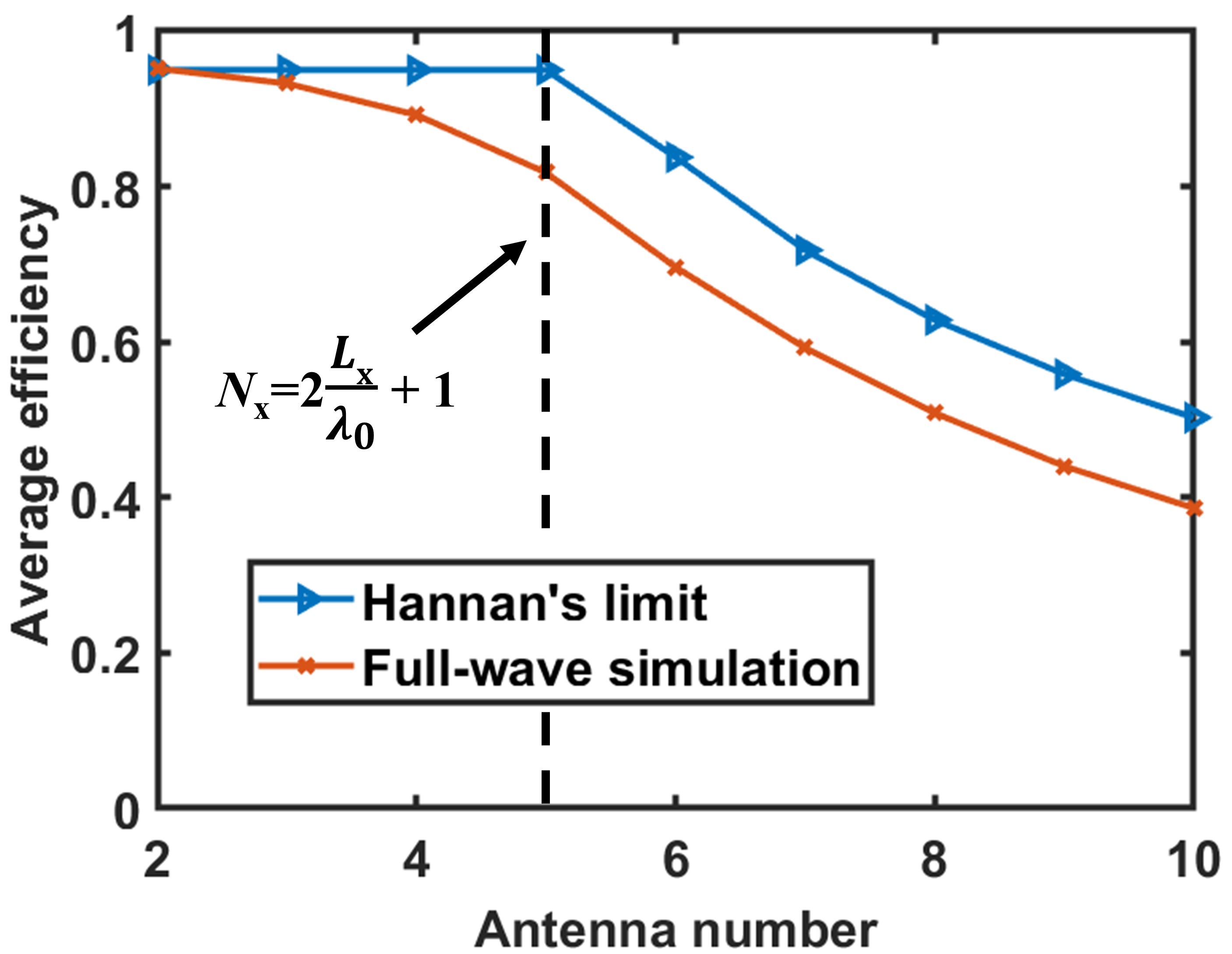}
		\caption{Average embedded radiation efficiencies obtained from full-wave simulations and Hannan's limit, and the efficiencies in Hannan's limit are regarded as 0.95 when antenna number $N_x<2L_x/\lambda_0+1$.}
		\label{MIMO}
	\end{figure}
	When the mutual coupling is ignored, we can observe that the diversities calculated with isotropic radiation patterns are slightly larger than that calculated with isolated patterns, because the non-isotropic radiation pattern of an isolated antenna is equivalent to a smaller angular spread. 
	\subsection{EFFECT OF MUTUAL COUPLING ON RADIATION EFFICIENCY}
	The embedded radiation efficiencies obtained from simulated $S$ parameters are depicted in Fig. 7. As expected, the efficiencies will drop quickly when the antenna number is increasing, especially for the antennas around the central region (power will be coupled to nearby antennas). Notice that the working frequency of the antenna array will shift to high frequencies due to mutual coupling, the errors brought by this frequency shift should be taken into account when the bandwidths of the array elements are narrow. Then, we compare the average efficiencies obtained from full-wave simulations and Hannan's limit, as shown in Fig. 8. 
	\begin{figure}[ht!]
		\centering
		\includegraphics[width=3in]{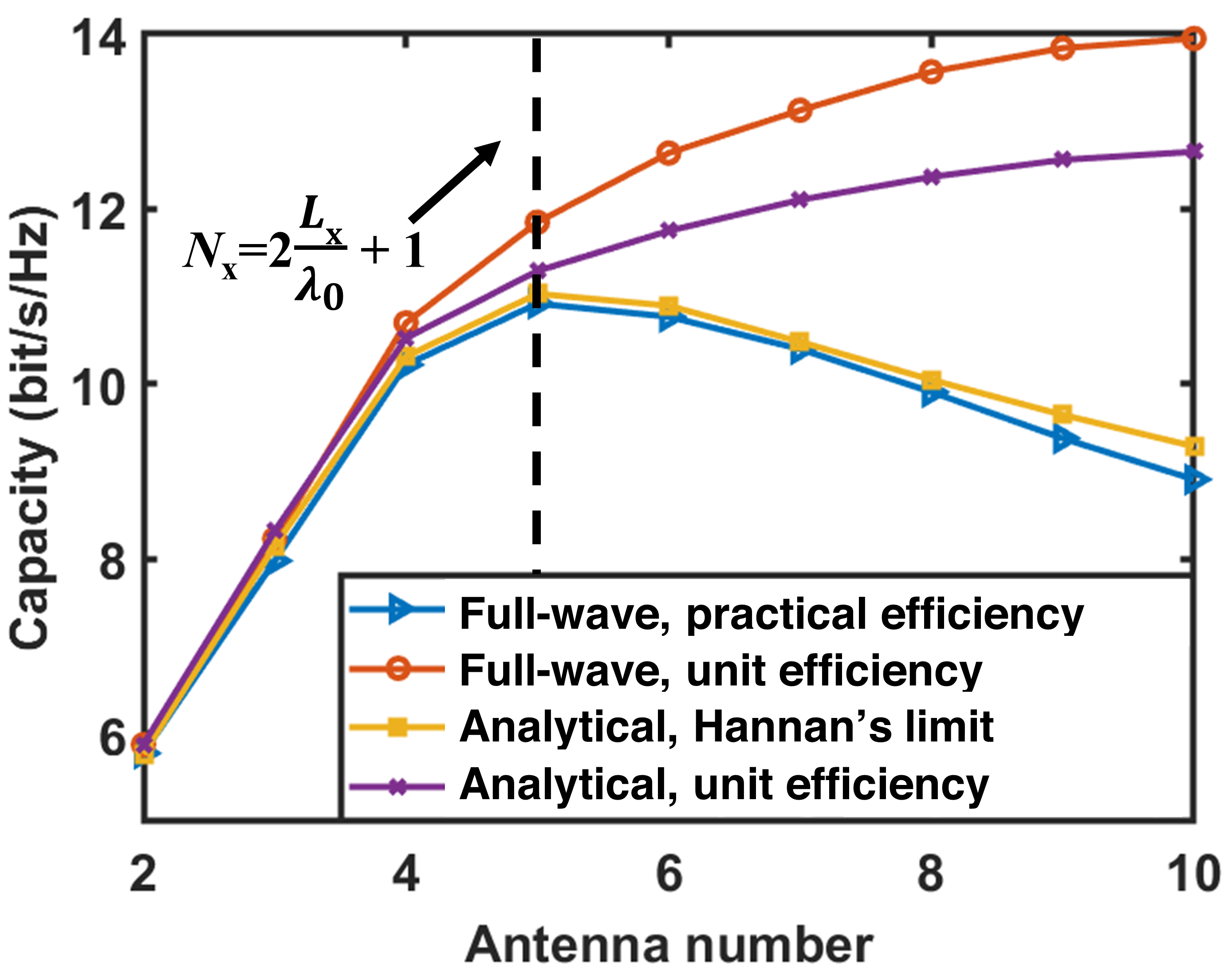}
		\caption{{Capacities of the aperture-constrained MIMO systems in different scenarios, total SNR $\gamma = 10$ dB.}}
		\label{MIMO}
	\end{figure}
	Regarding Hannan's limit, the average antenna efficiency is taken as 0.95 when the antenna number is smaller than $2L_x/\lambda_0+1$, assuming the antennas can maintain relatively high efficiencies at large element spacing. It can be found from Fig. 8 that the decreasing tendencies of the efficiencies are almost the same at small element spacing ($N_x>2L_x/\lambda_0+1$), and the efficiencies given by Hannan's limit actually indicate an upper bound. Hence, it would be effective and convenient to apply Hannan's limit to model the decrease of efficiency in holographic MIMO communications.
	\subsection{EFFECT OF MUTUAL COUPLING ON CAPACITY}
	{According to (6-16), the channel capacities in different scenarios are calculated based on both full-wave simulations and analytical methods. As shown in Fig. 9, the mutual coupling is generally detrimental to the capacity of a MIMO communication system.} When the element number becomes larger than $2L_x/\lambda_0+1$, the DOF will not further increase, however, the radiation efficiencies will keep decreasing, which results in a decreasing capacity. As a consequence, the holographic MIMO communications, or the tightly-coupled arrays, are hard to bring extra benefit compared to traditional uniform arrays with nearly $\lambda_0/2$ element spacing. Although the working bandwidth could increase through the tightly-coupling technology \cite{Volakis2013,Gao2019,Guo2018}, the dramatic decrease of efficiency would be fatal for MIMO applications. 
	
	We can also find the reasonable applicability of Hannan's limit at small element spacing. In Fig. 9, the capacities incorporating the Hannan's limit without coupling are almost the same as the practical full-wave simulation results. Although the efficiencies given by Hannan's limit are slightly higher than the practical values, the DOF without coupling is also slightly smaller than the practical DOF. Consequently, synthesizing the analytical expressions of DOF and Hannan's efficiency would obtain a capacity pretty close to the practical value. Therefore, in the channel modeling of holographic MIMO communications, (5) can be readily embedded for accurately characterizing the antenna effects at small element spacing. 
	\section{CONCLUSION}
	In this paper, we investigate how the mutual coupling affects the capacity of holographic MIMO system from the perspectives of DOF and radiation efficiency. The principles of correlation, DOF and efficiency are discussed with detailed analytical models, then numerical examples based on printed dipole arrays are illustrated to compare the practical MIMO performances with the analytical models. In the presented numerical results, the strong mutual coupling will slightly enhance DOF, but dramatically decrease antenna efficiency at small element spacing. Synthesizing the influences of coupling on DOF and efficiency will finally yield a reduced capacity when the element spacing is smaller than nearly half wavelength. Even if ideal decouplings are made, the DOF limit of a space-constrained MIMO system cannot be broken with planar antenna arrays. {Future explorations for breaking the DOF limit may consider new array architectures to better match both spatial and spectral characteristics of incoming waves in different communication channels. The new architectures include multi-layered antenna arrays with optimized spatial topology and reconfigurable antenna arrays with tunable angular spread. }

\vspace{6pt}


%


	
	
	\bibliographystyle{unsrt}  
	\bibliography{Bibliography}
\end{document}